\pdfoutput=1
\documentclass[draft]{agujournal2025}
\usepackage[strict]{changepage}
\usepackage{amsmath}

\makeatletter
\@ifundefined{fullwidth}{%
  \newlength{\draftfullwidthoverhang}%
  \setlength{\draftfullwidthoverhang}{5.75pc}%
  \newenvironment{fullwidth}%
    {\begin{adjustwidth}{-\draftfullwidthoverhang}{-\draftfullwidthoverhang}%
     \setlength{\linenumbersep}{\dimexpr\draftfullwidthoverhang+10pt\relax}}%
    {\end{adjustwidth}}%
}{}
\makeatother

\newcommand{\includeplannedfigure}[1]{%
  \includegraphics[width=0.97\linewidth]{#1}%
}

\begin{document}

\journalname{Geophysical Research Letters}

\title{Tropical Cyclone Attributes in Sparse Autoencoder Features of an AI Weather Model}

\authors{Jiapei Liu\affil{1}, Ruidong Li\affil{1}, Ting Sun\affil{2}, Guangheng Ni\affil{1}}

\affiliation{1}{State Key Laboratory of Hydro-science and Engineering, Department of Hydraulic Engineering, Tsinghua University, Beijing 100084, China}
\affiliation{2}{Department of Risk and Disaster Reduction, University College London, London WC1E 6BT, UK}

\authoraddr{Corresponding author: Ting Sun, Department of Risk and Disaster Reduction, University College London, London WC1E 6BT, UK. (ting.sun@ucl.ac.uk)}
% The draft class displays correspondence through title-page footnotes.
\ifpub\else
\correspondingauthor{Ting Sun}{ting.sun@ucl.ac.uk}
\fi

\authorrunninghead{Liu et al.}
\titlerunninghead{Cyclone Attributes in SAE Features of an AI Weather Model}

\keypoints
{Each attribute relates to multiple sparse autoencoder features; position and pressure-structure relations overlap in a smaller subset.}
{This organization recurs across independent sparse decompositions and all GraphCast processor layers.}
{In deeper layers, leading position features better match cyclone position and pressure structure six hours ahead than at present.}

\maketitle

\begin{abstract}
Artificial intelligence (AI) weather models now forecast weather events skillfully, but how they represent these events internally remains unknown.
Sparse autoencoders (SAEs) provide sparse features for probing hidden states, but how information about multiple physical attributes of an event is organized across them remains unclear.
Using tropical cyclones in GraphCast as a test case, we analyze 2,115 six-hourly timesteps from 205 cyclones, relating SAE activations to position, intensity, and pressure structure across five Layer-8 decompositions and 16 processor layers.
Each attribute relates to multiple features; position and pressure-structure relations overlap in a smaller subset.
This organization recurs across independent SAEs and layers without feature alignment.
In deeper layers, the leading position feature aligns better with position and pressure structure six hours ahead than at present.
These results suggest that the organization of physical-attribute relations provides a reproducible level of description for interpreting cyclone-related information in GraphCast.

  \begin{plainlanguagesummary}
  AI weather models can forecast weather events well, but this does not tell us how a weather event is represented inside these models.
We took cyclones as a case study.
A cyclone has several related properties, including its location, intensity, and the spatial structure of its pressure field.
We represented GraphCast’s internal states as combinations of many simpler components and asked how these properties were organized across them.
These properties were neither tied to one unique component nor cleanly separated into different components.
Instead, each property related to several components, while location and pressure structure shared some of those components.
Although the exact components varied across independent trainings, the same overall organization reappeared.
The same organization was also visible across GraphCast’s processing layers.
We then held each layer’s best location-matching component fixed and compared it with the cyclone’s current state and its state six hours later.
For location and pressure structure, components selected from earlier layers matched the current state more closely, whereas those selected from deeper layers matched the later state more closely.
Together, these results suggest that how cyclone properties are organized offers a reproducible description of cyclone information in GraphCast.
  \end{plainlanguagesummary}
\end{abstract}

\section{Introduction}

Artificial intelligence (AI) weather models have recently achieved global medium-range forecast skill competitive with conventional numerical weather prediction, and in February 2025 the European Centre for Medium-Range Weather Forecasts (ECMWF) made its Artificial Intelligence Forecasting System (AIFS) operational alongside its physics-based model \cite{lang2024aifs,moldovan2026aifs}.
These models predict complex atmospheric evolution at substantially lower computational cost and show strong skill for high-impact events such as tropical cyclones \cite{bi2023accurate,lam2023learning,price2024probabilistic,bodnar2025foundation}.
These forecasts are produced through black-box computations over high-dimensional hidden states inside these models, which cannot be directly interpreted in physical terms.
A weather event can be characterized by multiple measurable physical attributes that describe different aspects of the same evolving system \cite{casas2023intensity}.
These attributes provide physical reference points for interpreting event-related information in these hidden states.
Using multiple physical attributes to probe event-related information in the model therefore provides a physically grounded route to understanding how weather events are internally represented and to moving from predictive capability toward scientific understanding of these models.

Pursuing this aim requires resolving high-dimensional hidden states into components that can be related to these physical attributes.
Interpretability research in weather and climate science has largely focused on model input--output behavior, relating predictions to influential input variables or spatial regions \cite{mcgovern2019blackbox,toms2020physically,yang2024interpretable}, and therefore does not directly reveal how physical information is organized within hidden states.
Recent work has begun to examine processor states directly, for example by decoding intermediate GraphCast layers back into meteorological fields \cite{beylich2026intermediate}.
Sparse autoencoders (SAEs) take model hidden states as input and reconstruct them as sparse weighted combinations of many learned directions, with each direction defining a feature and its weight giving the corresponding feature activation \cite{huben2024sparse}.
The resulting features provide individually analyzable components through which internal signals associated with different physical information can be identified and compared more selectively.
SAEs have increasingly been used to examine internal representations in geophysical machine-learning models \cite{king2025sparse} and AI weather models \cite{macmillan2025mechanistic,mathai2026atmospheric}.
For tropical cyclones specifically, previous studies have identified SAE features whose activations track cyclone position \cite{macmillan2025mechanistic,cheon2026beyond}; intervention on a cyclone-related feature has also produced measurable changes in the forecast fields \cite{macmillan2025mechanistic}.
For atmospheric rivers, GraphCast SAE features track intensity as an internal variable and persist across processor depth \cite{mathai2026atmospheric}.
These studies establish that cyclone-related physical information can be identified through SAE features, but leave unresolved how relations to the multiple physical attributes of the same event are organized across these features.
Meanwhile, SAE features depend on the learned decomposition, and independently trained SAEs need not recover the same set of features \cite{leask2025canonical,paulo2025sparse}, although seed-dependent features can still span reproducible subspaces \cite{gerasimov2026unstable}.
Existing GraphCast SAE analyses rely on a single SAE fit per layer.
A further question is therefore whether this multi-attribute organization recurs across independently learned SAE decompositions.
As a weather model processes recent atmospheric states through successive layers to predict a future state, we also ask whether these attribute relations remain recognizable across depth and how their correspondence to current and forecast states changes.

Within this framework, we use SAEs to examine how the multiple physical attributes of a coherent weather event are organized across learned SAE features in an AI weather model, taking tropical cyclones in GraphCast as the test case.
Cyclones' identifiable footprints provide spatial targets, while their lifecycles allow relations between SAE features and multiple measurable physical attributes to be tested for recurrence through time.
We first test whether relations between physical attributes and SAE features recur consistently across time and event evolution (Section~\ref{sec:temporal-recurrence}).
At a fixed reference layer of GraphCast, we then determine how position, intensity, and pressure structure are shared and distributed across the full SAE dictionary (Section~\ref{sec:layer8-organization}).
We next test the reproducibility of this organization across independently trained SAE decompositions (Section~\ref{sec:layer8-organization}), and finally examine how the organization and its physical correspondence change along model depth (Section~\ref{sec:depth-organization}).

\section{Methods and Data}

\subsection{GraphCast and Sparse Autoencoders}

GraphCast is a graph-neural-network global weather model that uses atmospheric states at \(t-6\)~h and \(t\) to predict the state at \(t+6\)~h \cite{lam2023learning}.
It encodes gridded states onto a spherical mesh, updates them through 16 processor layers, and decodes them back to the latitude--longitude grid.
We use the \(0.25^\circ\) checkpoint trained on ERA5, the fifth-generation ECMWF global reanalysis, from 1979 to 2017 \cite{hersbach2020era5}.
At each processor layer, we analyze the 512-dimensional hidden states at 40,962 mesh nodes.
At each mesh node \(n\) and time \(t\), TopK SAEs encode the 512-dimensional hidden state \(\mathbf h(n,t)\) into 4,096 nonnegative coefficients, retaining \(K=32\) nonzero activations.
The decoder reconstructs the hidden state as
\[
\hat{\mathbf h}(n,t)=\mathbf b_{\mathrm{dec}}+\sum_{k=1}^{4096} z_k(n,t)\mathbf d_k,
\]
where the learned decoder directions \(\{\mathbf d_k\}\) form the SAE dictionary and \(z_k(n,t)\) is the activation of feature \(k\).
Each dictionary direction defines a feature; when its activation is compared with an external physical quantity, we refer to the feature as a readout.
With only 32 of 4,096 activations nonzero at each node, the features form a tractable set of candidate readouts, whose physical correspondence is evaluated below.
Architecture and optimization (Text~S1 and Table~S1) follow the previous GraphCast SAE setup \cite{macmillan2025mechanistic}.
The SAEs are trained on 1979--2017 GraphCast hidden states and validated for reconstruction on 2018--2021 states (Text~S1).
The cyclone analysis uses only the 2018--2019 GraphCast states.
For detailed dictionary-level analysis, we use Layer 8, near the midpoint of GraphCast's 16 processor layers, as a fixed intermediate depth reference, and subsequently extend the analysis across all processor layers.

\subsection{Cyclone and Pressure References}\label{sec:cyclone-reference}

Cyclone locations and spatial extents are obtained from the International Best Track Archive for Climate Stewardship (IBTrACS) \cite{knapp2010international,gahtan2024ibtracs}.
For each valid 2018--2019 record, the reported cyclone centre and 34-kt wind radius (R34) define its spatial footprint and corresponding cyclone target.
When multiple cyclones occur simultaneously, their footprints are merged into one timestamp-level target.
The record contains 2,115 cyclone-valid six-hourly timestamps, 4,118 storm--time records, and 205 cyclones.

Mean sea level pressure (MSLP) is taken from ERA5 \cite{hersbach2020era5} and paired with GraphCast states at six-hourly resolution.
For each target calendar day and UTC hour, we calculate a smoothed daily climatology from 1979--2019 same-UTC samples within \(\pm30\) calendar days \cite{jung2008scale}.
Samples are weighted by

\begin{equation}
w(d)=1-\left(\frac{d}{31}\right)^2,
\end{equation}

where \(d\) is the calendar-day distance from the target date, wrapped cyclically across year boundaries.
The \(\pm30\)-day window retains a similar seasonal background while the taper downweights more distant dates.
The MSLP anomaly is defined as

\begin{equation}
p'(n,t)=p(n,t)-\overline{p}_{\mathrm{clim}}(n,t),
\end{equation}

where negative values indicate lower pressure relative to the local seasonal background.
The deepest negative MSLP anomaly within the cyclone serves as a pressure-based intensity proxy, whereas the spatial anomaly field provides the structure reference.

\subsection{Cyclone--Feature Correspondence Measures}\label{sec:correspondence-measures}

We quantify position, intensity, and spatial pressure structure every 6 h over 2018--2019.
Position uses the merged timestamp-level target, whereas intensity and structure are evaluated first for individual storm--time cases.

Position correspondence is measured using F1, the harmonic mean of precision and recall between the feature's active nodes and the cyclone mask.
Let \(A_k(t)=\{n\colon z_k(n,t)>0\}\) denote the positive support of feature \(k\), and let \(M(t)\) denote the cyclone target at the same time.
Their spatial overlap is

\begin{equation}
F_{1,k}(t)= \frac{2\lvert A_k(t)\cap M(t)\rvert}{\lvert A_k(t)\rvert+\lvert M(t)\rvert}.
\end{equation}

F1 accounts for both cyclone coverage and concentration of feature activation within the cyclone.
Each feature's positional relation is summarized by median F1 across cyclone-valid timestamps.

Intensity correspondence is measured using the signed deviation \(D\).
For cyclone \(s\) at time \(t\), with footprint \(M_s(t)\), physical intensity is represented by the deepest anomalous low pressure within the footprint:

\begin{equation}
I^{\mathrm{phys}}_{s,t}=-\min_{n\in M_s(t)}p'(n,t),
\end{equation}

and the response amplitude of feature \(k\) is represented by its maximum activation within the same footprint:

\begin{equation}
I^{\mathrm{feat}}_{k,s,t}=\max_{n\in M_s(t)}z_k(n,t).
\end{equation}

The two quantities are separately min--max normalized over all 2018--2019 storm--time records.
Physical intensity uses one common range, whereas each feature uses its own activation range.
We then define

\begin{equation}
D_{k,s,t}=\widetilde{I}^{\mathrm{feat}}_{k,s,t}-\widetilde{I}^{\mathrm{phys}}_{s,t}.
\end{equation}

Values of \(D\) near zero indicate similar normalized amplitudes, with the sign indicating a relatively stronger or weaker feature response.
For simultaneous cyclones, storm-level \(D\) values are averaged to obtain \(D_k(t)\), so deviations of opposite sign can partly cancel.
The full-period relation is characterized by its temporal distribution, with \(\operatorname{median}_t D_k(t)\) as a directional summary.

Spatial pressure structure is measured using storm-centred Spearman correlation.
For each storm--time case, we compare feature activation with the spatial structure of the MSLP anomaly within a local \(\pm12^\circ\) region centred on the cyclone.
The calculation domain is the union of the cyclone target and the positive support of the feature within this region.
We calculate the Spearman correlation between continuous feature activation and the reversed MSLP anomaly, \(-p'\):

\begin{equation}
\rho_{k,s}(t)=\operatorname{Spearman}\left(z_k(n,t),-p'(n,t)\right).
\end{equation}

Positive \(\rho\) indicates closer spatial rank correspondence between feature activation and the reversed MSLP-anomaly structure.
For simultaneous cyclones, we use the median valid storm-level correlation.
Each feature is summarized by the distribution and temporal median.

\subsection{Quantifying Cyclone-Attribute Organization and Its Evolution Across Model Depth}\label{sec:organization-analysis}

To quantify how broadly each cyclone-attribute relation is distributed across SAE features, we convert the full-period summaries to nonnegative favorable-direction scores:
\begin{equation}
x_i^{F1}=F_{1,i},\qquad
x_i^{D}=\max\left(0,1-\frac{L_i}{L_0}\right),\qquad
x_i^{\rho}=\max(\rho_i,0).
\end{equation}

Here, \(L_i=\operatorname{median}_t|D_i(t)|\), and \(L_0=\operatorname{median}_t|D_0(t)|\) is the corresponding error of a zero-activation readout:
\begin{equation}
D_0(t)=-\frac{1}{|\mathcal S_t|}\sum_{s\in\mathcal S_t}\widetilde I^{\mathrm{phys}}_{s,t},
\end{equation}
where \(\mathcal S_t\) is the set of cyclones present at time \(t\).
The common physical-intensity reference over the shared evaluation timestamps gives \(L_0\simeq0.252\).
We call \(x_i^D\) D-proximity, which quantifies improvement in median absolute intensity deviation relative to a zero-activation readout.
A value of 1 indicates zero median absolute deviation, while 0 indicates no improvement over the zero-activation baseline or worse-than-baseline agreement.
Within each fixed evaluation set \(\mathcal G\), we normalize each attribute separately and define its effective feature number:
\begin{equation}
p_i^a=\frac{x_i^a}{\sum_{j\in\mathcal G}x_j^a},\qquad
N_{\mathrm{eff}}^a=\frac{1}{\sum_{i\in\mathcal G}(p_i^a)^2},
\qquad
a\in\{F1,D,\rho\}.
\end{equation}

Zero-weight features remain in \(\mathcal G\); these quantities are defined when \(\sum_{j\in\mathcal G}x_j^a>0\).
Larger \(N_{\mathrm{eff}}^a\) indicates broader feature participation.
We define the F1--pressure-structure overlap \(O\) by
\begin{equation}
\begin{aligned}
O&=\sum_i\min\left(p_i^{F1},p_i^{\rho}\right),\\
s_i&=\frac{\min\left(p_i^{F1},p_i^{\rho}\right)}{O}, \qquad O>0,\\
N_{\mathrm{shared}}&=\frac{1}{\sum_i s_i^2}.
\end{aligned}
\end{equation}
Here, \(\min(p_i^{F1},p_i^\rho)\) are the overlap weights, and \(s_i\) are the normalized overlap weights.
\(N_{\mathrm{shared}}\) is the effective feature number of the overlap.

For the Layer-8 reference analysis, dictionary-wide quantities are calculated over all features with jointly finite summaries.
For detailed multi-attribute comparison, all 4,096 features within each SAE dictionary are ranked by contemporaneous median F1, and the four highest-ranked features are compared using F1, signed \(D\), and pressure-structure correlation (Section~\ref{sec:layer8-organization}).
The Top~4 size was determined from the seed-50108 Layer-8 dictionary and then held fixed across all 80 seed--layer dictionaries.

Layer 8 is the primary analysis layer, for which five SAEs are trained independently with different random seeds.
The same F1 ranking and multi-attribute comparison are performed separately within each SAE dictionary.
For the cross-layer analysis, we use the same five random seeds, 50108--50112, at each of GraphCast's 16 processor layers, with seed 50108 providing the main layer-wise profile and the five-seed family providing the corresponding across-fit comparison.
Feature selection is performed independently within each seed--layer dictionary, and no feature alignment is imposed across SAE fits or processor layers.

For the forecast-time comparison, each seed--layer dictionary selects its own contemporaneous F1-leading feature.
That feature and its activation at time \(t\) are held fixed while the physical reference is changed from \(t\) to \(t+6\)~h.
Position F1, signed \(D\), and pressure-structure correlation are calculated against both physical states, with differences defined as \(t+6\) minus \(t\).
The five fit-specific differences are summarized at each matched time and processor layer by their median.
This comparison therefore measures how the physical-time alignment of the fixed readout at each layer changes across processor depth, without aligning feature identities across fits or layers.

\section{Results}

\subsection{Temporal Recurrence and Variability of Cyclone Attribute--Feature Relations}\label{sec:temporal-recurrence}

\begin{figure*}[p]
  \begin{fullwidth}
  \centering
  \includeplannedfigure{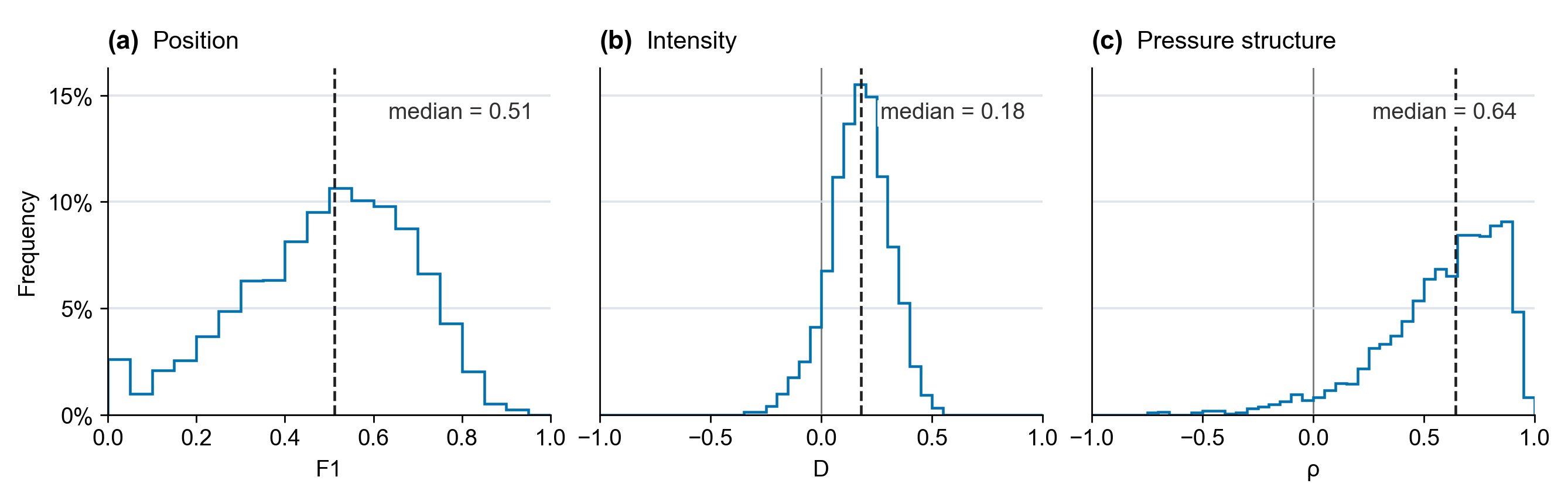}
  \caption{Temporal distributions of cyclone-attribute relations for GraphCast Layer-8 feature \texttt{f613} in 2018--2019: (a) position F1, (b) signed intensity deviation \(D\), and (c) pressure-structure correlation.}
  \label{fig:attribute-stability}
  \end{fullwidth}
\end{figure*}

We first tested whether cyclone attribute--feature relations recurred across the 2018--2019 record.
In the reference Layer-8 SAE, the contemporaneous F1 leader, defined as the feature with the highest median F1, was \texttt{f613}, where 613 is its identifier in the 4,096-feature SAE dictionary.
Across 2,115 cyclone-valid timestamps, median position F1, signed \(D\), and pressure-structure correlation were 0.512, 0.180, and 0.644, respectively (Figure~\ref{fig:attribute-stability}).
This position correspondence is comparable to that of the cyclone-tracking feature previously reported for the same Layer-8 SAE configuration \cite{macmillan2025mechanistic}, so the reference SAE reproduces that single-fit result, which the following analyses extend.
The recurring position and pressure-structure correspondence identifies \texttt{f613} as a cyclone-related readout, while substantial temporal variation in all three measures motivates using temporal medians for subsequent feature comparisons.

\begin{figure*}[p]
  \begin{fullwidth}
  \centering
  \includeplannedfigure{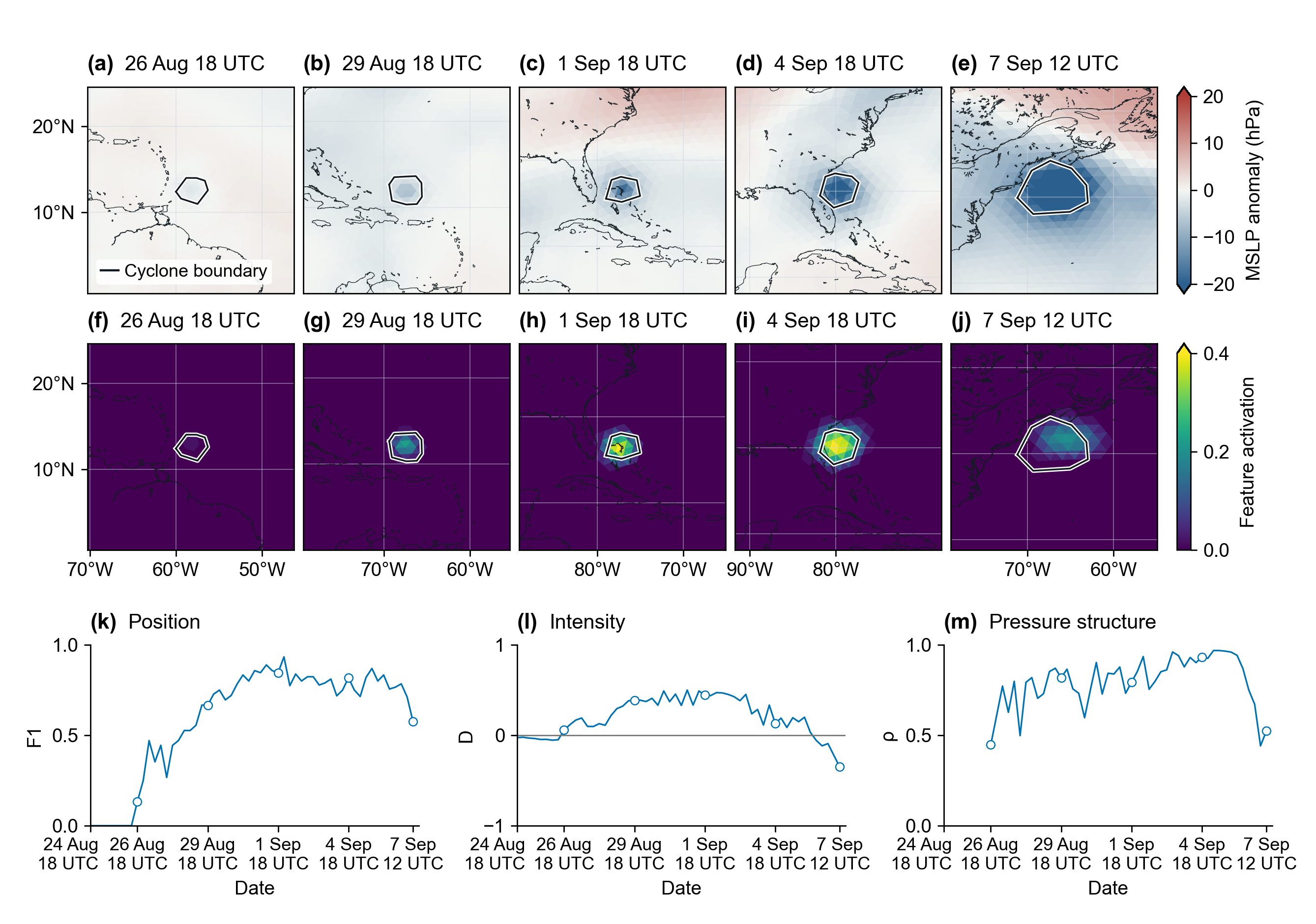}
  \caption{Dorian lifecycle in GraphCast Layer-8 feature \texttt{f613} in 2019. (a--j) Storm-centred maps of MSLP anomaly (a--e) and corresponding feature activation (f--j) at the five dates and UTC times indicated. (k--m) Six-hourly trajectories of position F1, signed intensity deviation \(D\), and pressure-structure correlation from 24 August at 18 UTC to 7 September at 12 UTC; open circles mark the displayed map times. Outer map ticks show absolute longitude and latitude.}
  \label{fig:dorian-lifecycle}
  \end{fullwidth}
\end{figure*}

To illustrate how the relations of \texttt{f613} to cyclone attributes vary through a storm's lifecycle, we examine North Atlantic Hurricane Dorian, a long-lived, well-tracked cyclone that reached exceptional intensity, over 24 August--7 September 2019, spanning its developing, mature, and weakening stages (Figure~\ref{fig:dorian-lifecycle}).
Early in this period, the cyclone target was present while \texttt{f613} activation remained weak and spatially incomplete, producing low F1 and weak or undefined pressure-structure correlation.
As Dorian organized and intensified, activation concentrated around the cyclone; at 1 September 2019, 18 UTC, F1 reached 0.846 and pressure-structure correlation 0.794.
Near the end, activation weakened while the physical target persisted, and both relations declined.
Signed \(D\) remained generally positive during the mature stage and became negative toward the end.

\subsection{Cyclone-Attribute Organization Across SAE Features}\label{sec:layer8-organization}

\begin{figure*}[p]
  \begin{fullwidth}
  \centering
  \includeplannedfigure{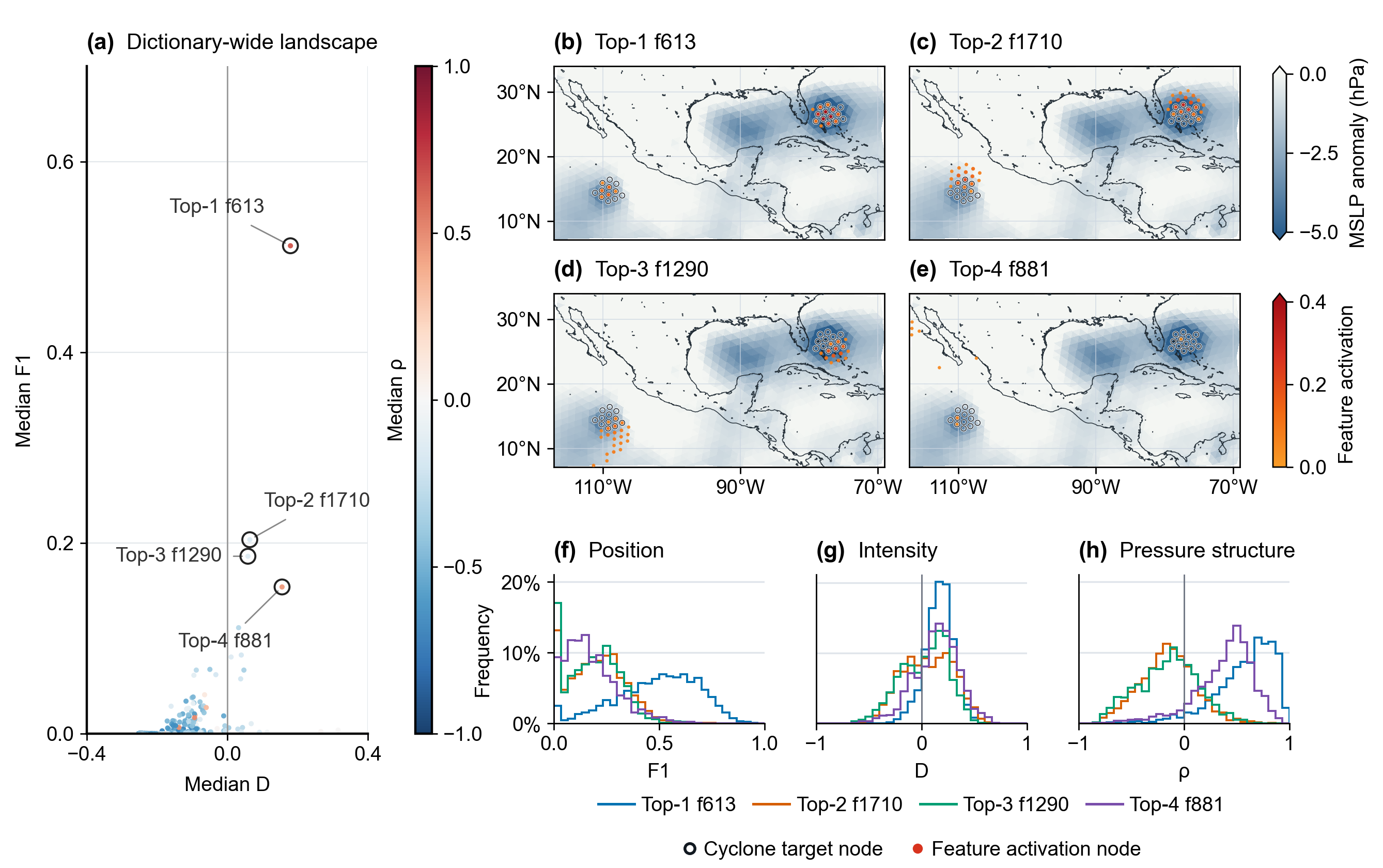}
  \caption{Layer-8 organization of cyclone-attribute relations. (a) Dictionary-wide landscape for the 3,589 features with finite summaries for all three measures, with the F1-selected Top~4 highlighted. Colour represents the temporal median pressure-structure correlation \(\rho\). Feature labels (e.g., \texttt{f613}) denote SAE feature IDs. (b--e) Spatial patterns of the Top~4 on the common 1 September 2019, 18 UTC Dorian--Juliette field. (f--h) Complete 2018--2019 distributions of position F1, signed intensity deviation \(D\), and pressure-structure correlation.}
  \label{fig:layer8-organization}
  \end{fullwidth}
\end{figure*}

Having established temporal recurrence, we next moved from the reference feature to the dictionary-wide organization of cyclone-attribute relations.
The Layer-8 SAE contains 4,096 features, of which 3,589 yielded finite full-period summaries for all three attribute measures (Figure~\ref{fig:layer8-organization}a).
To characterize how each cyclone-attribute relation was distributed across SAE features, we calculated dictionary-wide effective feature numbers of \(N_{\mathrm{eff}}^{F1}=18.5\), \(N_{\mathrm{eff}}^{D}=703.1\), and \(N_{\mathrm{eff}}^{\rho}=59.5\).
Position and pressure-structure correspondence were distributed across restricted but non-singleton subsets of the dictionary, whereas \(D\)-proximity was distributed much more broadly.
Position correspondence was highly uneven: most features had median F1 below 0.1, whereas \texttt{f613} led at 0.512, followed by 0.203, 0.186, and 0.154 for the next three features.

The dictionary-wide F1--pressure-structure overlap weights were more concentrated than either individual attribute distribution, with \(N_{\mathrm{shared}}=4.38\).
To examine how multiple cyclone-attribute relations coexist among leading position-related SAE features, we selected the F1-ranked Top~4 (\texttt{f613}, \texttt{f1710}, \texttt{f1290}, and \texttt{f881}; Figure~\ref{fig:layer8-organization}b--h).
This size was chosen exploratorily after inspecting the reference dictionary-wide landscape, as the smallest F1-ranked subset containing multiple features with positive pressure-structure correlation alongside others with negative correlation.
In the representative field containing Dorian and Juliette, all four showed cyclone-related activation, but their spatial patterns differed in displacement, extent, and completeness.
Across the full record, \texttt{f613} combined the highest position F1 with the strongest pressure-structure correlation (0.644), and \texttt{f881} also showed positive correlation (0.442), whereas \texttt{f1710} and \texttt{f1290} showed negative correlations.
Thus, positive pressure-structure correspondence accompanied strong position correspondence in some, but not all, of the selected features.

Together, the dictionary-wide measures and selected-feature comparisons reveal a distributed but partially bundled organization: cyclone-attribute relations extend across multiple SAE features, while position and pressure-structure relations overlap in a more concentrated subset.

To test whether this organization depended on one particular decomposition, we repeated the Layer-8 analysis across five independently trained SAEs.
Within each fit's F1-selected Top~4, \(N_{\mathrm{eff}}=1\) indicates concentration in one feature and \(N_{\mathrm{eff}}=4\) equal participation across all four.
Across fits, effective feature numbers ranged from 3.08--3.94 for position F1, 2.57--3.97 for \(D\)-proximity, 1.16--2.88 for positive pressure-structure correlation, and 1.39--2.88 for the F1--pressure-structure overlap (Table~S2).
Thus, F1 was distributed across about 3--4 features and \(D\)-proximity across about 2--4 features, whereas feature participation in pressure-structure relations and in the overlap ranged from strong concentration to moderate distribution across about 1--3 features.
The leading feature's temporal-median position F1 ranged from 0.345 to 0.512 and its margin over the second-ranked feature from 0.027 to 0.308; in seed 50112, the F1 leader (\texttt{f277}) was not the Top~4 feature with the strongest pressure-structure correlation (\texttt{f3984}; Text~S2, Figure~S1).
This recurring pattern reproduced the distributed but partially bundled organization across SAE decompositions despite changes in feature identity and ranking.

\subsection{Organization and Physical-Time Alignment Across Processor Depth}\label{sec:depth-organization}

\begin{figure*}[p]
  \begin{fullwidth}
  \centering
  \includeplannedfigure{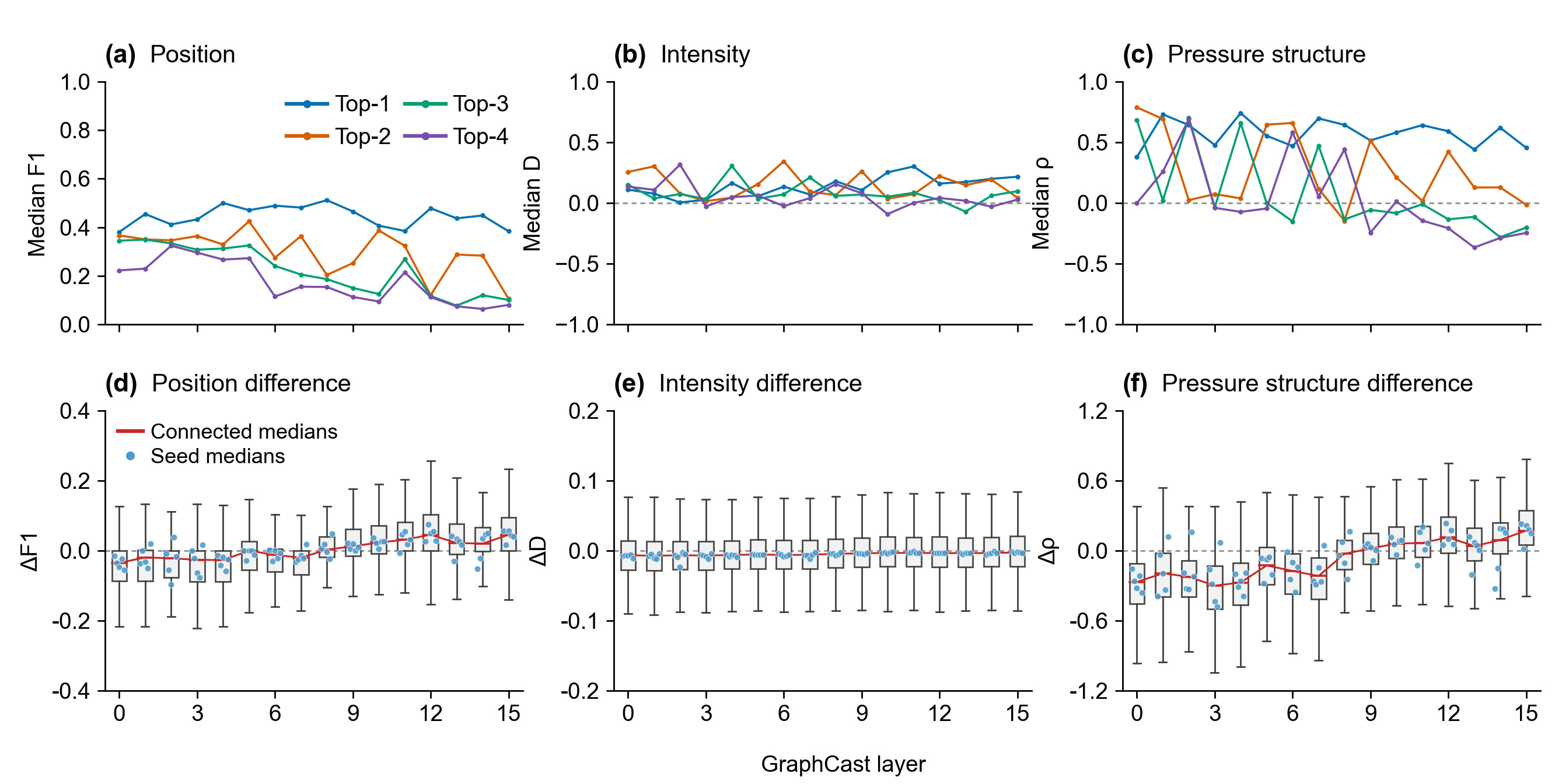}
  \caption{Organization and physical-time alignment across GraphCast processor depth. (a--c) Layer-wise temporal medians of position F1, signed intensity deviation \(D\), and finite-only pressure-structure correlation for the F1-selected Top~4 of SAE seed 50108. (d--f) Five-seed fixed-feature \(t+6\)-minus-\(t\) differences in the same three measures for contemporaneously selected F1-leading features. Grey boxes summarize the temporal distribution of the median difference across five seeds at each matched time. Boxes span the interquartile range; whiskers extend to the outermost observed values within 1.5 interquartile ranges. Red lines connect temporal medians of these differences, and blue points show each seed's temporal median. For correlation boxes and connected medians, only matched times with at least three finite seed values are used.}
  \label{fig:depth-organization-alignment}
  \end{fullwidth}
\end{figure*}

We then followed this organization through processor depth, as the underlying GraphCast representation changed from layer to layer.
Across all 16 layers of the reference seed (50108), each layer's F1 leader retained positive median pressure-structure correlation (0.379--0.741) and was among the top two features by correlation in 13 of 16 layers (Figure~\ref{fig:depth-organization-alignment}a--c).
Across the other four SAE fits, F1-leading features also showed positive median pressure-structure correlation at every processor layer.
The Top~1--Top~2 F1 gap and the correlation ordering within the F1-selected Top~4 varied across seeds and layers (Figure~S9).
Across seeds and layers, lower-F1 features sometimes showed higher pressure-structure correlation or median signed \(D\) closer to zero than the F1 leader.
Effective feature numbers within each F1-selected Top~4 showed the same pattern across depth: position and \(D\)-proximity involved multiple leading readouts, while feature participation in pressure-structure relations and in the overlap ranged from single-feature concentration to more distributed configurations (Table~S3).
Thus, despite changes in feature identity, ranking, and concentration across layers, the distributed but partially bundled organization remained detectable throughout processor depth.
Persistence across depth has also been reported for atmospheric-river SAE features \cite{mathai2026atmospheric}; here the comparison covers all 16 processor layers, with five independent SAE fits at each.

Finally, we tested whether the readouts selected at each layer remained aligned with the same physical time.
For each of five independently trained SAEs at every layer, we fixed its contemporaneous F1-leading feature and compared the same activation at time \(t\) against physical references at \(t\) and \(t+6\)~h.
After taking the median across the five fit-specific differences at each matched time, the temporal median of \(\Delta\mathrm{F1}\) was nonpositive through Layer 7 and positive from Layer 8 onward, while that of \(\Delta\rho\) was negative through Layer 8 and positive from Layer 9 onward (Figure~\ref{fig:depth-organization-alignment}d--f).
The median of signed \(\Delta D\) remained close to zero across layers.
The clearest depth-dependent change was therefore that deeper layers' F1-leading features aligned better with cyclone position and pressure structure six hours ahead than at present.
This event-level shift complements field-level evidence that forecasts decoded from intermediate GraphCast states improve gradually through the processor \cite{beylich2026intermediate}.

\section{Discussion and Conclusions}

Using tropical cyclones in GraphCast as the test case, we identify a distributed but partially bundled organization of event-attribute relations through SAE readouts.
Relations to individual attributes extend across multiple features, while position and pressure-structure relations overlap in a smaller subset.
The observed relations are neither confined to one feature nor cleanly separated among attribute-specific features.
This organization recurs across independent SAEs and layers.
In deeper layers, the F1-leading features align better with cyclone position and pressure structure six hours ahead than at present, while the median change in signed intensity deviation remains close to zero.

The recurrence of this organization across independently trained SAEs suggests that event-related information should not be identified with one fixed SAE feature.
Independent SAEs can express the same model hidden states in different sparse coordinates.
These differences may involve feature splitting, feature merging, and other forms of non-uniqueness in sparse decomposition, consistent with evidence that learned SAE dictionaries need not recover canonical units \cite{leask2025canonical}, that features differ across training seeds, particularly for TopK SAEs \cite{paulo2025sparse}, and that seed-dependent features can concentrate in reproducible subspaces \cite{gerasimov2026unstable}.
This study therefore compares the organization of attribute relations among each fit's F1-selected features without aligning individual features across fits, and finds that this organization recurs.

The observed organization may reflect three sources: covariation among the physical attributes of the event itself, how this coupled information is encoded in GraphCast hidden states, and the particular sparse decomposition learned by the SAE.
Our analysis characterizes the organization observable through SAE readouts but does not distinguish the contributions of these sources.
SAE dictionary size, sparsity, and other training choices may affect feature splitting and aggregation, thereby altering the observed distribution and overlap of attribute relations.
We did not test sensitivity to these SAE settings or to the F1-selected subset size, on which the within-subset \(N_{\mathrm{eff}}^\rho\) and \(N_{\mathrm{shared}}\) in Tables~S2 and S3 depend.
The analysis also starts from predefined physical attributes—position, intensity, and pressure structure—and therefore examines a selected physical description of the event.
A complementary approach would start from model-derived features or subspaces and ask which event dimensions emerge from the representation itself.
Our analysis is also confined to one model, GraphCast, and to two years (2018--2019, 205 cyclones) of single-step states; other architectures, longer samples, and autoregressive rollouts may organize these relations differently.
Finally, the readouts establish correspondence rather than use: whether GraphCast relies on these features for its forecasts would require intervention or ablation experiments \cite{macmillan2025mechanistic}.

Because the SAEs at each layer are fitted independently, they reveal this organization at different depths through separate SAE features, without directly tracing the transfer of event-related information between layers.
Aligning features or subspaces across layers could help test how this information is carried and updated through the processor.

Extending the framework to other weather phenomena requires phenomenon-specific physical definitions: the event and its measurable attributes must first be specified before their relations to sparse readouts can be evaluated.
Tropical cyclones provide a particularly well-defined case because their position, life cycle, and multiple physical attributes can be tracked consistently.
Fronts, atmospheric rivers, convection, and other less compact phenomena may require different physical measures and may exhibit different organizational patterns.
Testing these phenomena would help distinguish cyclone-specific structure from more general properties of weather-event representations.

These findings suggest that, in the cyclone case, the organization of physically meaningful attribute relations provides a reproducible target for interpreting event-related information in GraphCast's hidden states.

\section*{Acknowledgments}

The authors acknowledge the use of resources provided by the Isambard-AI National AI Research Resource (AIRR) under project u6lp \cite{mcintoshsmith2024isambardai}.
Isambard-AI is operated by the University of Bristol and supported by the UK Government's Department for Science, Innovation and Technology (DSIT) through UK Research and Innovation and the Science and Technology Facilities Council under award ST/AIRR/I-A-I/1023.
J.~Liu was supported by the Tsinghua University Visiting Doctoral Students Foundation (Contract No.~2025146) for his research visit to University College London.

\section*{Conflict of Interest}

The authors declare there are no conflicts of interest for this manuscript.

\section*{Open Research}

GraphCast source code and pretrained model resources are available through the official WeatherNext repository \cite{deepmind2026weathernext}.
We used GraphCast at commit \texttt{a64f793}.
ERA5 fields \cite{hersbach2020era5} used as GraphCast inputs and MSLP references were obtained from the WeatherBench~2 archive \cite{rasp2024weatherbench2} (\url{https://storage.googleapis.com/weatherbench2/datasets/era5/1959-2022-full_37-6h-0p25deg_derived.zarr}).
Tropical cyclone tracks and wind-radius data were obtained from IBTrACS v4.01 \cite{gahtan2024ibtracs}.
SAE checkpoints, analysis code, and processed results are archived at Zenodo \cite{liu2026cyclonearchive} (\url{https://doi.org/10.5281/zenodo.23492703}).

\bibliography{references}

\end{document}